\documentclass[11pt,twoside]{article}
\usepackage{asp2004}
\usepackage{psfig}
\usepackage{epsf}
\usepackage{graphics}
\usepackage{lscape}
\def\edcomment#1{\iffalse\marginpar{\raggedright\sl#1\/}\else\relax\fi}
\reversemarginpar

\begin{document}
\title{First Results from THINGS: The HI Nearby Galaxy Survey}
\author{W.J.G. de Blok}
\affil{School of Physics and Astronomy, Cardiff University, PO Box 913, Cardiff CF24 3YB, UK}
\author{F. Walter}
\affil{NRAO, P.O. Box O, Socorro, NM 87801, USA}
\author{E. Brinks}
\affil{INAOE, Apdo. Postal 51 \& 216, Puebla, Pue 72000, Mexico}
\author{M.D. Thornley}
\affil{Dept. of Physics, Bucknell University, Lewisburg, PA 17837, USA}
\author{R.C. Kennicutt, Jr.}
\affil{Steward Observatory, University of Arizona, Tucson, AZ 85721, USA}

\begin{abstract}
We describe \emph{The HI Nearby Galaxy Survey (THINGS)}, the largest program
ever undertaken at the VLA to perform 21-cm HI observations of the
highest quality ($\sim 7''$, $\leq 5$ km\, s$^{-1}$ resolution) of
nearby galaxies. The goal of THINGS is to investigate key
characteristics related to galaxy morphology, star formation and mass
distribution across the Hubble sequence.  A sample of 34 objects with
distances between 3 and 10 Mpc will be observed, covering a wide range
of evolutionary stages and properties. Data from THINGS will
complement SINGS, the Spitzer Infrared Nearby Galaxy Survey.  For the
THINGS sample, high-quality observations at comparable resolution will
thus be available from the X-ray regime through to the radio part of
the spectrum. THINGS data can be used to investigate issues such as
the small-scale structure of the ISM, its three-dimensional structure,
the (dark) matter distribution and processes leading to star
formation.  To demonstrate the quality of the THINGS data products, we
present some prelimary HI maps here of four galaxies from the THINGS
sample.
\end{abstract}
\thispagestyle{plain}

\section{Introduction}

Studies of the atomic interstellar medium (ISM), through observations
of the 21-cm line of atomic hydrogen (HI), have proven to be critical
for our understanding of the processes leading to star formation, the
dynamics and structure of the ISM, and the (dark) matter distribution;
thus touching on major issues related to galaxy formation.  In the
last decades remarkable progress has been made in these areas.
However, the lack of high-resolution HI observations in a
representative sample of nearbly galaxies precludes a systematic study
of the physical characteristics of the atomic ISM.

Here we present first maps of THINGS, `The HI Nearby Galaxy Survey', the
largest program ever undertaken at the VLA to perform 21-cm atomic hydrogen
observations of the highest quality of nearby galaxies.

The goal of THINGS is to obtain high-quality observations of the
atomic ISM (through observations of the 21-cm line of neutral atomic
hydrogen) of a substantial sample of nearby galaxies, covering a wide
range of Hubble types, star formation rates, absolute luminosities,
evolutionary stages, and metallicities.  The database will have
homogeneous sensitivity and spatial and velocity resolution at the
limit of what can currently be achieved with the Very Large Array. The
angular resolution will be $\leq 7''$, and the spectral resolution
will be $\leq 5$ km s$^{-1}$. All galaxies will be observed with the
VLA B, C and D arrays.  A more complete description of THINGS, survey
parameters and observational and technical details will be presented
in Walter et al.\ \citetext{in preparation}. A list of sample targets
is given in Table 1. As an example of the high quality of the THINGS data,
Figs.\ 1 and 2 show high-resolution integrated HI column density maps
of four galaxies from the sample.

Data from THINGS will complement SINGS, the Spitzer Infrared Nearby
Galaxies Survey \citep{sings}; hence high-quality observations from
the X-ray through the radio will be available at comparable resolution
for each galaxy. Data from THINGS can be used to investigate issues
such as the small-scale and three-dimensional structure of the ISM,
the (dark) matter distribution, and the processes leading to star
formation. Furthermore, THINGS will enable studies of the variation of
each of these properties as a function of galaxy environment.

\section{Scientific Goals}

THINGS is designed to address numerous scientific topics, including, but not
limited to, the following:
\begin{itemize}
\item {\bf Dust -- Atomic ISM.} The resolution of the THINGS HI data is similar to that obtained for the intermediate wavelength bands on-board Spitzer. One
of the main goals of THINGS is therefore a comparison between the spatial distributions
of the HI and dust as measured by Spitzer.

\item {\bf Interplay ISM -- Star Formation.} THINGS allows studies of the
interplay between star formation and the ambient ISM at 100-300 pc resolution
over a range of Hubble types. The location and energy input of regions of
recent star formation and the impact they have on the structure and dynamics
of the HI will be investigated. For example, for the first time a census will be possible
of supergiant shells as a function of Hubble type. THINGS data products
will enable studies of how these structures form and how they, in
turn, might trigger secondary star formation. Using the
multi--wavelength data from SINGS a complete energy budget of the ISM
can be derived.  

\item {\bf Global Mass Distribution.} The THINGS 
data products will enable studies of the dark matter distribution in
galaxies at high angular resolution. High resolution data near the
centres of galaxies are especially important as this is the regime
where differences between the 'cuspy' and 'constant-density' haloes
show themselves most clearly. The data from Spitzer and additional
data at other wavelengths will deliver more reliable estimates for the
stellar mass--to--light ratio of the disk, obviating the need to
resort to the ``canonical'' maximum disk assumption in order to derive
the properties of the dark matter halo.  This study will ultimately
shed more light on the validity of the CDM paradigm on scales of
individual galaxies and will measure possible trends of the dark
matter properties as a function of Hubble type.

\item {\bf Star Formation Threshold.} Using THINGS, studies will be performed
at high spatial resolution to investigate whether or not there is a
`universal' star formation threshold and how, or even if, this is a
function of galaxy type. Combinations of THINGS HI surface density
maps with CO data will allow one to create total gas surface density
maps. Amongst other things, with the velocity dispersion maps and the
derived rotation curves, this information can be used to calculate
spatially resolved ``Toomre--$Q$'' parameter maps for each galaxy (the
$Q$ parameter is a measure for the local gravitational balance,
\citealt{toomre}) as a function of Hubble type. Issues that can be
addressed are, e.g., in which regime does `$Q$' break down? What
is the importance of local (disk or cloud instability) versus global
effects (spiral density waves, tidal forces)?

\end{itemize}

\section{THINGS data products}

Obervations for THINGS are currently ongoing at the VLA (total observing time: $\sim 500$ hours, including archival data). The data acquisition will be completed at the end of 2005. Final data products of THINGS will include moment maps (integrated HI distribution, velocity fields, velocity dispersion) as well
as the HI data cubes, and will be made available to the community one year 
after the observations have completed.

\begin{table}[!ht]
\caption{Summary of THINGS targets}
\smallskip
\begin{center}
{\footnotesize
\begin{tabular}{lllllrrcrrrr}
\tableline
\noalign{\smallskip}
Name       & Alias    & Type & RA (2000.0) & dec (2000.0)& V$_{\rm rad}$ & W$_{20}$ & size & incl \\
           &          &      & h  m s.s   &$^{\circ}$ $'$ $''$&km\,s$^{-1}$&km\,s$^{-1}$& $'\times'$&  $^{\circ}$ \\
\tableline
\noalign{\smallskip}
%IC10       &          & Irr  & 00 20 17.3 & +59 18 14 &--348& 80  &XX&XX\\
NGC628     & M74      & Sc   & 01 36 41.7 & +15 46 59 & 656 &  77 & 10.5 $\times$ ~9.5 & 35    \\
NGC925     &          & SBcd & 02 27 16.9 & +33 34 45 & 553 & 217 & 10.5 $\times$ ~5.9 & 61    \\
NGC1569    &          & IBm  & 04 30 49.0 & +64 50 53 & --90&  83 & ~3.6 $\times$ ~1.8 & 68   \\
NGC2366    &          & Irr  & 07 28 47.6 & +69 11 39 &  99 & 114 & ~8.1 $\times$ ~3.3 & 90   \\
NGC2403    &          & SBc  & 07 36 51.4 & +65 36 09 & 130 & 241 & 21.9 $\times$ 12.3 & 60   \\
HolmbergII & UGC4305  & Irr  & 08 19 04.0 & +70 43 09 & 157 &  72 & ~7.9 $\times$ ~6.3 & 47   \\
M81dwA     & PGC23521 & Irr  & 08 23 56.0 & +71 01 45 & 113 &  33 & ~1.3 $\times$ ~0.7 & --  \\
DDO53      & UGC4459  & Irr  & 08 34 07.2 & +66 10 54 &  19 &     & ~1.5 $\times$ ~1.3 & 45   \\
%He2--10    & PGC24171 & E--SO& 08 36 15.0 &--26 24 34 & 873 & 155 & ~1.9 $\times$ ~1.4 & 48   \\
NGC2841    &          & Sb   & 09 22 02.6 & +50 58 35 & 638 & 607 & ~8.1 $\times$ ~3.5 & 68   \\
NGC2903    &          & SBbc & 09 32 10.1 & +21 30 04 & 555 & 384 & 12.6 $\times$ ~6.0 & 56   \\
HolmbergI  & UGC5139  & Irr  & 09 40 32.3 & +71 10 56 & 137 &  44 & ~3.6 $\times$ ~3.0 & 37   \\
NGC2976    &          & Sc   & 09 47 15.3 & +67 55 00 &   3 & 135 & ~5.9 $\times$ ~2.7 & 61   \\
NGC3031    & M81      & Sab  & 09 55 33.2 & +69 03 55 &--35 & 442 & 26.9 $\times$ 14.1 & 59   \\
NGC3077    &          & Sd   & 10 03 20.6 & +68 44 04 &  13 &  90 & ~5.4 $\times$ ~4.5 & 41   \\
M81dwB     & UGC5423  & Irr  & 10 05 30.6 & +70 21 52 & 343 &  61 & ~0.9 $\times$ ~0.6 & 67   \\
NGC3184    &          & SBc  & 10 18 16.9 & +41 25 28 & 591 & 151 & ~7.4 $\times$ ~6.9 & 24   \\
NGC3198    &          & SBc  & 10 19 54.9 & +45 32 59 & 662 & 321 & ~8.5 $\times$ ~3.3 & 70    \\
IC2574     & UGC5666  & SBm  & 10 28 21.2 & +68 24 43 &  48 & 109 & 13.2 $\times$ ~5.4 & 75   \\
NGC3351    & M95      & SBb  & 10 43 57.8 & +11 42 14 & 778 & 280 & ~7.4 $\times$ ~5.0 & 42   \\
NGC3521    &          & SBbc & 11 05 48.6 &--00 02 09 & 809 & 462 & 11.0 $\times$ ~5.1 & 66    \\
NGC3621    &          & SBcd & 11 18 16.0 &--32 48 42 & 726 & 278 & 12.3 $\times$ ~7.1 & 66 \\
NGC3627    & M66      & SBb  & 11 20 15.0 & +12 59 30 & 726 & 377 & ~9.1 $\times$ ~4.2 & 57   \\
NGC4214    &          & Irr  & 12 15 38.9 & +36 19 40 & 290 &  86 & ~8.5 $\times$ ~6.6 & 42  \\
NGC4449    &          & Irr  & 12 28 11.2 & +44 05 36 & 202 & 143 & ~6.2 $\times$ ~4.4 & 56  \\
NGC4736    & M94      & Sab  & 12 50 53.0 & +41 07 14 & 309 & 232 & 11.2 $\times$ ~9.1 & 35   \\
DDO154     &          & Irr  & 12 54 05.2 & +27 08 55 & 373 &  93 & ~3.0 $\times$ ~2.2 & 44   \\
NGC4826    & M64      & Sab  & 12 56 43.7 & +21 40 52 & 406 & 315 & 10.0 $\times$ ~5.4 & 60   \\
NGC5055    & M63      & Sbc  & 13 15 49.2 & +42 01 49 & 502 & 400 & 12.6 $\times$ ~7.2 & 56   \\
NGC5194    & M51a     & Sbc  & 13 29 52.7 & +47 11 43 & 463 & 199 & 11.2 $\times$ ~6.9 & 30   \\
NGC5236    & M83      & SBc  & 13 37 00.8 &--29 51 59 & 515 & 281 & 12.9 $\times$ 11.5 & 46 \\
%NGC5253    &          & S?   & 13 39 55.9 &--31 38 24 & 402 &  92 & ~5.0 $\times$ ~1.9 & 67   \\
NGC5457    & M101     & SBc  & 14 03 12.5 & +54 20 55 & 241 & 188 & 28.8 $\times$ 26.9 & 22   \\
NGC6946    &          & SBc  & 20 34 52.3 & +60 09 14 &  51 & 235 & 11.5 $\times$ ~9.8 & 31   \\
NGC7331    &          & SAb  & 22 37 04.1 & +34 24 56 & 816 & 520 & 10.5 $\times$ ~3.7 & 71 \\
NGC7793    &          & Scd  & 23 57 49.7 &--32 35 30 & 229 & 192 & ~9.3 $\times$ ~6.3 & 53   \\
\noalign{\smallskip}
\tableline
\end{tabular}
}
\end{center}
\end{table}

\begin{figure}[!ht]
%\plotone{deblok_fig1.ps}
\caption{Integrated HI colum density maps of N3621 (top) and M81dwA (bottom).
Grayscale levels run from $1.0\cdot 10^{20}$ cm$^{-2}$ (light) to
$3.8\cdot 10^{21}$ cm$^{-2}$ (dark) for N3621, and from $1.0\cdot 10^{20}$ cm$^{-2}$ (light) to
$1.5\cdot 10^{21}$ cm$^{-2}$ (dark) for M81dwA. The beam is shown in the 
bottom-right corner}
\end{figure}

\begin{figure}[!ht]
%\plotone{deblok_fig2.ps}
\caption{Integrated HI colum density maps of Holmberg II (top) and M83 (bottom).
Grayscale levels run from $1.0\cdot 10^{20}$ cm$^{-2}$ (light) to
$4.3\cdot 10^{21}$ cm$^{-2}$ (dark) for Holmberg II, and from $1.0\cdot 10^{20}$ cm$^{-2}$ (light) to
$3.0\cdot 10^{21}$ cm$^{-2}$ (dark) for M83. The beam is shown in the bottom-right corner for Holmberg II and the top-right corner for M83.}
\end{figure}

\end{document}